\documentclass[preprint,showpacs,preprintnumbers,amsmath,amssymb]{revtex4}
\usepackage{graphicx}
\usepackage{dcolumn}
\usepackage{bm}
\usepackage[latin1]{inputenc} 
\newcommand{\comment}[1]{}

\begin{document}

\title{
Reduction of mean-square advection in turbulent passive scalar mixing
}

\author{Wouter J.T. Bos,{$^1$} Robert Rubinstein,{$^2$} and Le Fang{$^{1,3}$}}
\affiliation{$^1$ LMFA, CNRS, Ecole Centrale de Lyon -\\ Universit\'e de Lyon, Ecully, France\\
$^2$ Newport News, VA, USA\\
$^3$ Ecole Centrale de P\'ekin, Laboratoire International Associ\'e, Beihang University, Beijing 100191, Beijing, China}

\begin{abstract}

Direct numerical simulation data shows that the variance of the
coupling term in passive scalar advection
by a random velocity field is smaller than it would be if the velocity and scalar fields were
statistically independent. This effect is analogous to the 
`depression of nonlinearity' in hydrodynamic turbulence. 
We show that the
trends observed in the numerical data are qualitatively
consistent with the predictions of closure theories related to 
Kraichnan's Direct Interaction Approximation.
The phenomenon is demonstrated over a range of Prandtl numbers. In the inertial-convective range the depletion is approximately constant with respect to wavenumber. The effect is weaker in the Batchelor range.
\end{abstract}



\maketitle

\section{introduction}
The modal amplitudes in the Fourier decomposition of any homogeneous
random field are uncorrelated. In a Gaussian random field, they
are also statistically independent; but in homogeneous turbulence, 
nonlinearity produces statistical dependence among the amplitudes. 
The simplest consequence is that the
third-order correlations representing energy transfer, which would
vanish in a Gaussian random field, do not vanish in homogeneous turbulence.

Some further consequences of statistical dependence of Fourier amplitudes 
in homogeneous turbulence were considered
in an important paper by Chen, Herring, Kerr and Kraichnan,\cite{Chen1989}
which compared various fourth-order moments with the corresponding moments 
in a Gaussian random field with the same second-order properties as the turbulent velocity field 
(the construction of such
Gaussian surrogates is sometimes called `kinematic simulation' \cite{Fung1992,Kraichnan1970-2}). 
Among the quantities investigated by Chen \emph{et al.} was the variance of the
fluctuating nonlinear term in the Navier-Stokes equations,
\begin{equation}
\langle
\left|
{\bm u (\bm x,t)}\cdot\nabla{\bm u(\bm x,t)} + \nabla p(\bm x,t)
\right|^2
\rangle.
\label{nonl}
\end{equation}
It had been observed \cite{Kraichnan1988} that this quantity is significantly
smaller in a turbulent velocity field than in its Gaussian counterpart; 
that is, there is a significant (negative) cumulant contribution to the 
fourth order moment defined by Eq. (\ref{nonl}).
One of the mechanisms which can lead to this {\em depression of nonlinearity} is the preferential alignment of velocity and vorticity, also called {\em Beltramization}.\cite{Moffatt1992} However, this preferential alignment is not the only non-trivial mechanism which 
is consistent with the depression of nonlinearity;
we will return to this issue in Section \ref{sec:Discussion}.

From the viewpoint of a Fourier analysis of the spectrum of
the correlation in Eq. (\ref{nonl}),
the depression of nonlinearity is
a consequence of statistical dependence of the
uncorrelated Fourier amplitudes that enter the expression for this spectrum.
One finding of Chen \emph{et al.} was that
this phenomenon appears to be well predicted by Kraichnan's \cite{KraichnanDIA}
Direct Interaction Approximation (DIA). 
The successful prediction of a nonzero fourth-order cumulant by a closure
theory might seem unexpected or even surprising, since from the very beginning,
closure theories have been associated with cumulant discard hypotheses;\cite{Tatsumi1957}
the debate between Kraichnan and Proudman
at the famous 1961 Marseille conference \cite{Proudman1962} centered on this 
issue.\cite{Footnote1}
The computation of a nonzero cumulant and
the favorable comparison with data perhaps vindicate, somewhat after the fact,
Kraichnan's assertion at that time,\cite{Kraichnan1962} 
that DIA {\em does not assume (or imply)
the vanishing of fourth order cumulants.}

In the present work we will show that an effect of statistical
dependence of Fourier amplitudes analogous to depression of nonlinearity also
appears in 
the advection of a passive scalar $\theta$. 
Thus, we consider the scalar analog of the moment in Eq. (\ref{nonl}): the variance of 
fluctuations of the bilinear scalar-velocity coupling
\begin{equation}
\langle
({\bm u}({\bm x},t)\cdot\nabla\theta({\bm x},t))^2
\rangle.
\label{eq:adv}
\end{equation}
Herring and M\'etais \cite{Herring1992} have shown that 
this quantity is smaller in passive scalar advection than it would be if the 
Fourier amplitudes of velocity and scalar were statistically independent,
even at the more refined level of Fourier spectra.
We confirm their conclusions using higher resolution DNS data, and
following Chen \emph{et al.}, show that closures related to the DIA predict trends consistent
with the data.

A different perspective on non-Gaussian properties of turbulence is
provided by 
recent detailed studies of the properties of realizations of turbulent
velocity fields. Such studies,
made possible
by high resolution direct numerical simulations,\cite{Ishihara2009}
reveal the existence of flow structures such as 
vortex tubes and sheets, and spotty regions of very high dissipation; 
in comparison, since a Gaussian
random field is simply space- and time-filtered white noise, it is
expected to be essentially `featureless.' 
This viewpoint makes the existence of such structures
the most significant effect of non-Gaussianity in turbulence. 
In the present paper we focus on a statistical characterization of non-Gaussian features in turbulence and do not investigate features of the instantaneous flow realizations. We suggest, however, that investigating the connections between this physical space 
perspective and the viewpoint of dependence among Fourier modes can
be a useful direction for future research.

The paper is organized as follows: in Section \ref{sec:Analysis} the theoretical considerations leading to closure expressions for the mean-square advection term are given. 
Section \ref{sec:added} presents details of the numerical evaluation of cumulant 
corrections.
Section \ref{sec:Numerics} presents
comparisons between closure computations and direct numerical simulation data. Section \ref{sec:Discussion} contains a discussion of the results. Conclusions
are drawn in Section \ref{sec:conclusions}.

\section{analysis \label{sec:Analysis}}

We consider the advection of a passive scalar in homogeneous
turbulence. 
The governing equation is
\begin{equation}
\left[\partial_t+\alpha k^2\right]\theta(\bm k,t)=
-ik_i
\int d{\bm p}d{\bm q} \ 
\delta({\bm k-\bm p-\bm q})\theta({\bm p},t)u_i({\bm q},t)
+f_{\theta}(\bm k,t),
\label{equation}
\end{equation}
where $\alpha$ denotes the scalar diffusivity, 
and $f_{\theta}({\bm k},t)$ is a source of scalar fluctuations
that we will assume confined to the large scales. 
By analogy to Chen \emph{et al.},
we consider the contribution of each Fourier mode to the
variance of the velocity-scalar coupling term. It is defined by
\begin{eqnarray}
& &
W_{\theta}({\bm k},t) =
k_ik_j
\int d{\bm p}\ d{\bm q} \
\int d{\bm p}'\ d{\bm q}' \
\delta({\bm k-\bm p-\bm q})
\delta({\bm k+\bm p'+\bm q'})
\langle 
\theta({\bm p},t)u_i({\bm q},t)
\theta({\bm p}',t)u_j({\bm q}',t)
\rangle\nonumber\\
\label{W defn}
\end{eqnarray}
The integral of $W_{\theta}({\bm k},t)$ over all wavevectors is equal to the 
moment in Eq. (\ref{eq:adv}),
\begin{equation}
\int d{\bm k}\ W_\theta({\bm k},t)=\langle[\bm u({\bm x},t)\cdot \nabla \theta({\bm x},t)]^2\rangle.
\end{equation}
Without introducing any assumptions, we can write
\begin{eqnarray}
W_\theta({\bm k},t)=W_\theta^G({\bm k},t)+W_\theta^C({\bm k},t),
\label{G+C}
\end{eqnarray}
where $W_\theta^G({\bm k},t)$ is evaluated assuming the independence of the Fourier
amplitudes in Eq. (\ref{W defn}) and $W_{\theta}^C({\bm k},t)$ is
a cumulant correction. In the following we will consider the isotropic case. In that case the velocity and scalar are
uncorrelated. Then
\begin{equation}
W_\theta^G({\bm k},t)= 
k_ik_j
\int d{\bm p}\ d{\bm q} \
\delta({\bm k-\bm p-\bm q}) \
U_{\theta}({\bm p},t)U_{ij}({\bm q},t),
\label{WG}
\end{equation}
where
\begin{equation}
U_{ij}({\bm k},t) = 
\langle u_i({\bm k},t) u_j(-{\bm k},t) \rangle
\end{equation}
is the single-time velocity autocorrelation and
\begin{equation}
U_{\theta}({\bm k},t) = 
\langle \theta({\bm k},t) \theta(-{\bm k},t) \rangle
\end{equation}
is the single-time scalar autocorrelation.

We now analyze $W_{\theta}$ using Kraichnan's DIA theory.
There are many equivalent formulations of this theory, but for this analysis,
the Langevin model formulation \cite{Kraichnan1970} is the most convenient.
The DIA Langevin model for passive scalar advection replaces the exact governing equation 
Eq. (\ref{equation}) by
\begin{equation}
\left[\partial_t+\alpha k^2\right]\theta(\bm k,t)+
\int^t_0ds\ \eta_{\theta}(\bm k;t,s)\theta(\bm k,s) =
-ik_i \int d{\bm p}d{\bm q} \ 
\delta({\bm k-\bm p-\bm q})\xi_{\theta}({\bm p},t)\xi_i({\bm q},t)
+f_{\theta}(\bm k,t),
\label{langevin}
\end{equation}
where 
$\xi_{\theta}$ and $\xi_i$ are independent
Gaussian random variables with the same {\em two-time}
correlation functions as $\theta$ and $u_i$:
\begin{eqnarray}
& &
\langle \xi_{\theta}({\bm k},t) \xi_{\theta}(-{\bm k},t') \rangle =
\langle \theta({\bm k},t) \theta(-{\bm k},t') \rangle
= U_{\theta}({\bm k};t,t') 
\label{xi theta}\\& &
\langle \xi_i({\bm k},t) \xi_j(-{\bm k},t') \rangle =
\langle u_i({\bm k},t) u_j(-{\bm k},t') \rangle =
U_{ij}({\bm k};t,t') 
\label{xi u}
\end{eqnarray}
and the damping function $\eta_{\theta}$ is defined as
\begin{equation}
\eta_{\theta}({\bm k};t,t') = k_ik_j 
\int d{\bm p}\ d{\bm q} \ \delta({\bm k-\bm p-\bm q}) \
G_{\theta}({\bm p};t,t') U_{ij}({\bm q};t,t').
\label{eta defn}
\end{equation}
Here, $G_{\theta}$ is the {\em response function}, defined as the
inverse of the (formally) linear operator on the left side of
Eq. (\ref{langevin}). This linearity allows us to write, ignoring the contribution of the
scalar source term,
\begin{equation}
\theta({\bm k},t) = -ik_i\int^t_0 ds\ G_{\theta}({\bm k};t,s)
\int d{\bm p}d{\bm q} \ 
\delta({\bm k-\bm p-\bm q})\xi_{\theta}({\bm p},s)\xi_i({\bm q},s).
\label{G defn}
\end{equation}
This brings up an important feature of DIA, namely that it is not
closed in terms of the correlation function alone.
The introduction of the response function
is one major contribution of DIA to turbulence theory.\cite{Footnote2}
DIA provides equations of motion for both $G_{\theta}$ and
the correlation function $U_{\theta}$ related to the model Eq. (\ref{langevin}).
We refer to \cite{Newman1979} for details.

Paraphrasing Kraichnan's own description of DIA,
we see that it first replaces the nonlinear coupling by a random forcing
by surrogate statistically independent random fields with the same second-order
properties as the actual fields; this step suppresses any statistical dependence
among Fourier modes that develops under the exact evolution. These correlations
are then {\em modeled} by the damping provided by
$\eta_{\theta}$; then the transfer of scalar fluctuations
between modes is treated in DIA as the result of this damping
acting against the forcing. Perhaps the most important qualitative feature to note is that
the theory requires two-time statistics: this complication
is inevitable given that DIA attempts to describe
complex bilinear interactions by means of second-order statistics alone.


Thus, DIA can be described as the replacement
\begin{eqnarray}
-ik_i\int d{\bm p}d{\bm q} \ 
\delta({\bm k-\bm p-\bm q})\theta({\bm p},t)u_i({\bm q},t)
\rightarrow-\left[
\int^t_0ds\ \eta_{\theta}(\bm k;t,s)\theta(\bm k,s)\right. \nonumber\\
\left.+ik_i \int d{\bm p}d{\bm q} \ 
\delta({\bm k-\bm p-\bm q})\xi_{\theta}({\bm p},t)\xi_i({\bm q},t)\right],
\label{nl model}
\end{eqnarray}
where the arrow simply indicates modeling; at this point, there is no
assertion about an `approximation.'
Then the DIA model for the variance of the advection term is 
the variance of the right side of Eq. (\ref{nl model}):
\begin{eqnarray}
& &
\left\langle
\left|
\int^t_0ds\ \eta_{\theta}(\bm k;t,s) \theta(\bm k,s) 
+ik_i \int d{\bm p}d{\bm q} \ 
\delta({\bm k-\bm p-\bm q})\xi_{\theta}({\bm p},t)\xi_i({\bm q},t)
\right|^2
\right\rangle
\nonumber\\& & =
\int^t_0ds\int^t_0ds'\ \eta_{\theta}(\bm k;t,s) \eta_{\theta}(\bm k;t,s') 
\langle\theta(\bm k,s) \theta(-\bm k,s') \rangle
\label{term 1}\\& & -
2i k_i \int d{\bm p}d{\bm q} \ 
\delta({\bm k-\bm p-\bm q})
\int^t_0ds\ \eta_{\theta}(\bm k;t,s) 
\left\langle\xi_{\theta}(-{\bm p},t)\xi_i(-{\bm q},t) \theta(\bm k,s) \right\rangle
\label{term 2}\\& & +
k_ik_j \int d{\bm p}d{\bm q} \ \int d{\bm p}'d{\bm q}' \ 
\delta({\bm k-\bm p-\bm q}) \delta({\bm k-\bm p'-\bm q'})
\langle
\xi_{\theta}({\bm p},t)\xi_i({\bm q},t) \xi_{\theta}(-{\bm p}',t)\xi_j(-{\bm q}',t)
\rangle.
\label{term 3}
\end{eqnarray}

The rules for correlations of
Gaussian variables, and the relations Eqs. (\ref{xi theta}) and
(\ref{xi u}) give for the term in Eq. (\ref{term 3}),
\begin{eqnarray}
& &
k_ik_j \int d{\bm p}d{\bm q} \ \int d{\bm p}'d{\bm q}' \ 
\delta({\bm k-\bm p-\bm q}) \delta(-{\bm k-\bm p'-\bm q'})
\langle
\xi_{\theta}({\bm p},t)\xi_i({\bm q},t) \xi_{\theta}({\bm p}',t)\xi_i({\bm q}',t)
\rangle 
\nonumber\\& &=
k_ik_j \int d{\bm p}d{\bm q} \ \delta({\bm k-\bm p-\bm q}) 
U_{\theta}({\bm p},t)U_{ij}({\bm q},t) =
W_\theta^G({\bm k},t),
\end{eqnarray}
so that, as was evident from its definition, this term simply reproduces 
the Gaussian contribution Eq. (\ref{WG}).
The remaining terms are cumulant corrections. Obviously,
the term in Eq. (\ref{term 1}) is simply
\begin{eqnarray}
& &
\int^t_0ds\int^t_0ds'\ \eta_{\theta}(\bm k;t,s) \eta_{\theta}(\bm k;t,s') 
\langle\theta(\bm k,s) \theta(-\bm k,s') \rangle 
\nonumber\\& &= 
k_ik_j k_mk_n
\int^t_0ds\int^t_0ds'\ 
\int d{\bm p}\ d{\bm q} \ \delta({\bm k-\bm p-\bm q}) \
\int d{\bm p}'\ d{\bm q}' \ \delta({\bm k-\bm p'-\bm q'}) \
\times\nonumber\\& &
G_{\theta}({\bm p};t,s) U_{ij}({\bm q};t,s)
G_{\theta}({\bm p}';t,s') U_{mn}({\bm q}';t,s')
U_{\theta}({\bm k};s,s'),
\label{term 2 bis}
\end{eqnarray}
where we have used the definition Eq. (\ref{eta defn}) of $\eta_{\theta}$.

The term in Eq. (\ref{term 2}) is evaluated by expressing $\theta$ in terms of the
$\xi_{\theta}$ and $\xi_i$ using Eq. (\ref{G defn}), so that
\begin{eqnarray}
& &
-2i k_i \int d{\bm p}d{\bm q} \ 
\delta({\bm k-\bm p-\bm q})
\int^t_0ds\ \eta_{\theta}(\bm k;t,s) 
\left\langle\xi_{\theta}(-{\bm p},t)\xi_i(-{\bm q},t) \theta(\bm k,s) \right\rangle
\nonumber\\& & =
-2i k_i(-ik_j) \int d{\bm p}d{\bm q} \ 
\int d{\bm p}''d{\bm q}'' \ 
\delta({\bm k-\bm p-\bm q}) \delta({\bm k-\bm p''-\bm q''})
\int^t_0ds\ \int^s_0ds' \ 
\eta_{\theta}(\bm k;t,s) G_{\theta}({\bm k};s,s')
\times\nonumber\\& &
\left\langle\xi_{\theta}(-{\bm p},t)\xi_i(-{\bm q},t) 
\xi_{\theta}({\bm p}'',s')\xi_j({\bm q}'',s')
\right\rangle
\nonumber\\& & =
-2k_ik_jk_mk_n \int d{\bm p}d{\bm q} \ 
\int d{\bm p}'d{\bm q}' \ 
\delta({\bm k-\bm p-\bm q}) \delta({\bm k-\bm p'-\bm q'})
\int^t_0ds\ \int^s_0ds' \ 
\times\nonumber\\& &
G_{\theta}({\bm p}';t,s) U_{mn}({\bm q}';t,s)
G_{\theta}({\bm k};s,s')
U_{\theta}({\bm p};t,s')U_{ij}({\bm q};t,s').
\label{term 3 bis}
\end{eqnarray}

The cumulant contribution $W^C_{\theta}$ is the sum of the results
of Eqs. (\ref{term 2 bis}) and (\ref{term 3 bis}). But to express
the result in the most transparent form,
it will be useful to reformulate Eq. (\ref{term 2 bis}) somewhat: abbreviating
the integrand for simplicity,
\begin{eqnarray}
& &
\int^t_0ds\int^t_0ds'\ {\cal I}({\bm p,\bm q,\bm p',\bm q'};t,s,s') =
\left(\int^t_0ds\int^s_0ds'\ +\int^t_0ds\int^t_sds'\ \right)
{\cal I}({\bm p,\bm q,\bm p',\bm q'};t,s,s') 
\nonumber\\& &=
\left(\int^t_0ds\int^s_0ds'\ +\int^t_0ds'\int^{s'}_0ds\ \right)
{\cal I}({\bm p,\bm q,\bm p',\bm q'};t,s,s'), 
\end{eqnarray}
where the order of integration has been interchanged in the second term.
Since the integrand is invariant under the simultaneous interchanges of 
$s$, $s'$ and ${\bm p}$, ${\bm p}'$, we obviously have
\begin{eqnarray}
& &
 \int^t_0ds\int^t_0ds'\ {\cal I}({\bm p,\bm q,\bm p',\bm q'};t,s,s')  =
2\int^t_0ds\int^s_0ds'\ {\cal I}({\bm p,\bm q,\bm p',\bm q'};t,s,s'), 
\end{eqnarray}
so we can write
\begin{eqnarray}
& &
\int^t_0ds\int^t_0ds'\ \eta_{\theta}(\bm k;t,s) \eta_{\theta}(\bm k;t,s') 
\langle\theta(\bm k,s) \theta(-\bm k,s') \rangle 
\nonumber\\& &= 
2k_ik_j k_mk_n
\int^t_0ds\int^s_0ds'\ 
\int d{\bm p}\ d{\bm q} \ \delta({\bm k-\bm p-\bm q}) \
\int d{\bm p}'\ d{\bm q}' \ \delta({\bm k-\bm p'-\bm q'}) \
\times\nonumber\\& &
G_{\theta}({\bm p};t,s) U_{ij}({\bm q};t,s)
G_{\theta}({\bm p}';t,s') U_{mn}({\bm q}';t,s')
U_{\theta}({\bm k};s,s').
\label{term 2 bis bis}
\end{eqnarray}
Interchanging indices $(ij)$ and $(mn)$ and the wavevector arguments
$({\bm p,\bm q})$ and $({\bm p',\bm q'})$ and adding the result of
Eq. (\ref{term 3 bis}), we obtain
\begin{eqnarray}
& &
W^C_{\theta}({\bm k},t) = 
-2k_ik_jk_mk_n
\int d{\bm p}\ d{\bm q} \
\int d{\bm p}'\ d{\bm q}' \
\delta({\bm k-\bm p-\bm q})
\delta({\bm k-\bm p'-\bm q'}) \times 
\nonumber\\& &
\int^t_0ds\ \int^s_0ds'\ 
U_{ij}({\bm q};t,s) U_{mn}({\bm q}';t,s')
\times\nonumber\\& &
\left[
G_{\theta}({\bm p};t,s) 
G_{\theta}({\bm p}';t,s') 
U_{\theta}({\bm k};s,s')-
G_{\theta}({\bm p}';t,s) 
G_{\theta}({\bm k};s,s')
U_{\theta}({\bm p};t,s')
\right].
\label{Wtheta bis}
\end{eqnarray}
This expression makes clear an important property
of the DIA cumulant correction, namely that it vanishes identically,
independently of the velocity field,
in the scalar {\em non-diffusive truncated ensemble} 
when diffusivity $\alpha=0$ and a maximum wavenumber is imposed on the scalar fluctuations. 
This equilibrium ensemble is Gaussian, therefore all cumulants vanish.
The proof follows from the properties of this system, that the
scalar field is in steady-state equipartition, so that $U_{\theta}({\bm k},t)$ is
a constant, and the fluctuation-dissipation relation
\begin{equation}
U_{\theta}({\bm k};t,t') = 
U_{\theta}({\bm k}) [G_{\theta}({\bm k};t,t') +
G_{\theta}({\bm k};t',t)] 
\end{equation}
holds. (Note that the response function is {\em causal}:
$G_{\theta}({\bm k};t,t') = 0$ for $t'>t$.)
Substituting these relations in Eq. (\ref{Wtheta bis}) shows at once
that $W^C_{\theta}\equiv 0$ independently of the velocity field, as required.
We remark that this conclusion is a nontrivial check of
the DIA calculation, since DIA only treats moments, and the multipoint
probability density functions play no explicit role.

It is easily verified that
the same result holds for the cumulant corrections to the
mean-square nonlinearity in the analysis of the velocity field.\cite{Chen1989}

\section{numerical evaluation of the DIA cumulant corrections \label{sec:added}}

At this point, we introduce the assumption that the velocity
field is  time stationary and that
the scalar field is maintained in a steady state by a scalar source term.
Then numerical evaluation is greatly simplified by expressing the
results in terms of spectra rather than correlations.
If $W_{\theta}({\bm k})$ depends only on $k=|\bm k|$,
then the corresponding spectrum is
\begin{eqnarray}& &\label{defwt}
w_\theta(k)= \oint dS({\bm k})\ W_{\theta}({\bm k}) 
=4\pi k^2 W_{\theta}(k) 
\end{eqnarray}
and, corresponding to Eq. (\ref{G+C}), we have
\begin{eqnarray}
w_\theta(k)=w_\theta^G(k)+w_\theta^C(k).
\label{G+C bis}
\end{eqnarray}
We introduce the usual energy and scalar fluctuation spectra by
\begin{equation}
U_{ij}({\bm k}) = \frac{1}{4\pi k^2}E(k)(\delta_{ij}-k^{-2}k_ik_j)
\quad
U_{\theta}({\bm k}) = \frac{1}{2\pi k^2}E_{\theta}(k).
\end{equation}

With these simplifications, Eq. (\ref{WG}) can be reformulated, 
following procedures that are standard in the closure literature, as
\begin{eqnarray}
w_\theta^G(k)=k^3 \int_\Delta (1-z^2)E(p)E_\theta(q) \frac{dp}{p} \frac{dq}{q}.
\label{wG}
\end{eqnarray}
where, as usual,
the integration region $\Delta$ indicates that the wavenumbers
$k, p, q$ are the sides of a triangle 
and $z$ is the cosine of the angle between the sides of lengths $k$ and $p$.
The time integrations in Eq. (\ref{Wtheta bis}) are evaluated by
replacing the two-time
quantities by functions of time difference only, then passing to
the steady state limit $t\rightarrow\infty$. 
Since we wanted to be able to compute the cumulants under a variety of conditions,
we found it expedient to make the double time integrations of Eq. (\ref{Wtheta bis}) 
analytically computable by assuming simple exponential time-dependence
\begin{eqnarray}& &
G_{\theta}({\bm k};t-t') = e^{-\eta_{\theta}(k)(t-t')} H(t-t'),
\quad
U_{\theta}({\bm k};t-t') = U_{\theta}({\bm k})e^{-\eta_{\theta}(k)|t-t'|},
\nonumber\\& &
\quad
U_{ij}({\bm k};t-t') = U_{ij}({\bm k})e^{-\eta_{\theta}(k)|t-t'|}.
\label{exp}
\end{eqnarray}
As usual, $H(s)$ is the `Heaviside function' equal to one for $s>0$ and zero otherwise;
we have also introduced a `fluctuation-dissipation' relation in which the damping
function $\eta_{\theta}$ is the same in the scalar response function $G_{\theta}$
and two-time correlation function $U_{\theta}$.
The very commonly introduced exponential {\em ansatz} for the two-time dependence 
is also made by Herring and M\'etais;
we emphasize that we use it
entirely in the interest of analytical simplicity, and no assertion
is made that it approximates the two-time response that would actually be predicted
by DIA. But since two-time statistics enter our results only after integration over
all time-differences, any resulting errors are unlikely to be qualitatively important.

After making all of these simplifications, the cumulant spectrum is evaluated as
\begin{eqnarray}\label{eqwtC}
w_\theta^C(k)=\frac{1}{2}
\int_{\Delta}
\frac{dp}{p}\frac{dq}{q}
\int_{\Delta'}
\frac{dp'}{p'}\frac{dq'}{q'}
(1-z^2)kq^2E(p)(1-z'^2)kq'^2E(p')\times\nonumber\\
\left[
\left(\Xi_{kpqp'q'}+\Xi_{kp'q'pq}\right)E_\theta(k)-2\Xi_{kpqp'q'}(k/q)^{2}E_\theta(q)
\right],
\end{eqnarray}
where $\Delta'$ indicates that the wavenumbers
$k, p', q'$ are the sides of a triangle,
$z'$ is the cosine of the angle between the sides of lengths $k$ and $p'$, and the
time integrals yield
\begin{eqnarray}
\Xi_{kpqp'q'}= \frac{1}{\eta_{\theta}(k)+{\eta}(p')+{\eta}_{\theta}(q')}~\frac{1}{\eta(p)+\eta_{\theta}(q)+\eta(p')+\eta_{\theta}(q')}.
\end{eqnarray}

The spectra $E(k)$ and $E_{\theta}(k)$ are evaluated using EDQNM (Eddy-Damped Quasi-Normal Markovian) closures \cite{Orszag,Herring}
\begin{eqnarray}\label{eqEK}
\left[\frac{\partial }{\partial t}+2\nu k^2\right]E(k)=\int_{\Delta}\Theta(k,p,q)~[xy+z^3]pE(q)\left[k^2  E(p)-p^2E(k)\right]\frac{dp dq}{pq}+F(k)\\
\left[\frac{\partial }{\partial t}+2\alpha k^2\right]E_\theta(k)=\int_{\Delta}\Theta^\theta(k,p,q)~[1-y^2]kE(q)\left[k^2 E_\theta(p)-p^2 E_\theta(k)\right]\frac{dp dq}{pq}+F_\theta(k),\label{eqEKT}
\end{eqnarray}
in which $F(k)$ and $F_\theta(k)$ are external forcing terms confined to the smallest wavenumbers (Both $F(k)$ and $F_\theta(k)$ are unity for $k\le 2$ and zero elsewhere),  $x$ is the cosine of the angle between the sides of lengths $p$ and $q$, and $y$ is the cosine of the angle between the sides of lengths $k$ and $q$. 
The triad relaxation times $\Theta(k,p,q)$ and $\Theta^\theta(k,p,q)$ are
\begin{eqnarray}
\Theta=\frac{1}{\eta(k)+\eta(p)+\eta(q)},\qquad
\Theta^\theta=\frac{1}{\eta_\theta(k)+\eta(p)+\eta_{\theta}(q)}.
\end{eqnarray}
We use the (inverse) time-scales
\begin{equation}
\eta(k)=\lambda \sqrt{\int_0^k E(r)dr}+\nu k^2,\qquad 
\eta_\theta(k)=\lambda_\theta \sqrt{\int_0^k E(r)dr}+\alpha k^2  
\end{equation}
and we set the constants $\lambda=\lambda_\theta=0.5$. 
Note that $\eta$ and $\eta_\theta$ are the same quantities that appear in Eq. (\ref{exp}). An interesting perspective for future work would be the use of a Lagrangian two-time theory \cite{Kraichnan65,Kaneda81} or a self-consistent Markovian closure \cite{KraichnanTFM,Bos2006} to evaluate the cumulants, which would avoid the introduction of adjustable constants and {\it ad-hoc} formulation of damping time-scales. Computations are carried out on a logarithmically spaced grid with $36$ gridpoints per octave and results are evaluated when a steady state is reached.

\section{numerical comparisons \label{sec:Numerics}}

\begin{figure}
\includegraphics[width=.5\textwidth]{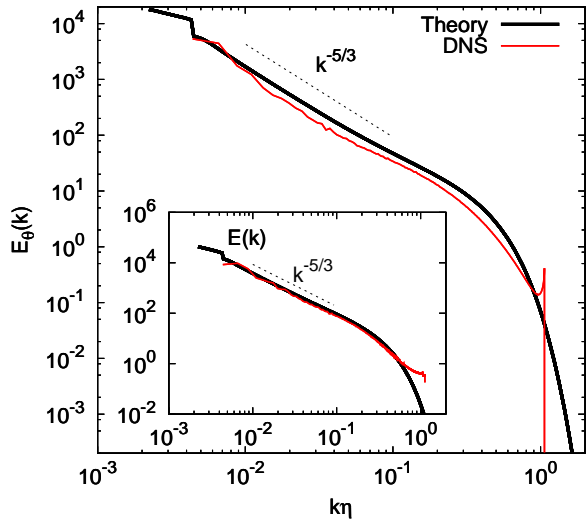}
\caption{DNS and theoretical results for the scalar spectra in isotropic turbulence at a Taylor-scale
 Reynolds number of $427$ and ${\mathcal Pr}=1$. In the inset the energy spectra are shown. 
\label{Fig1} }      
\end{figure}

\begin{figure}
\includegraphics[width=.4\textwidth]{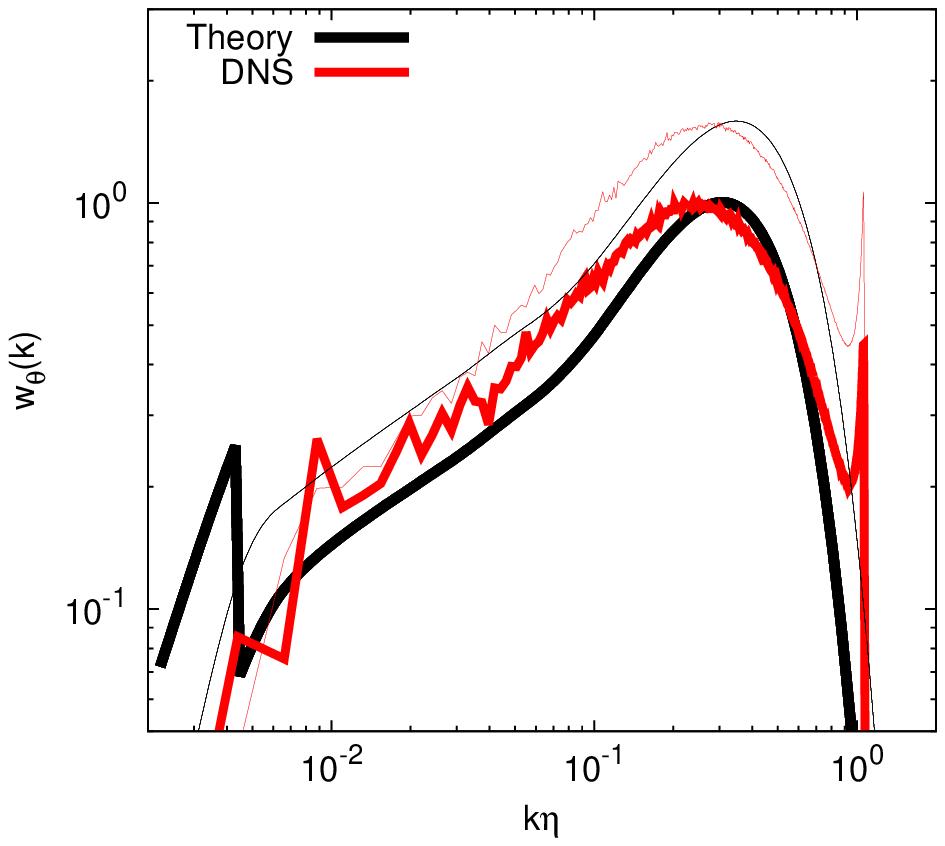}
\includegraphics[width=.4\textwidth]{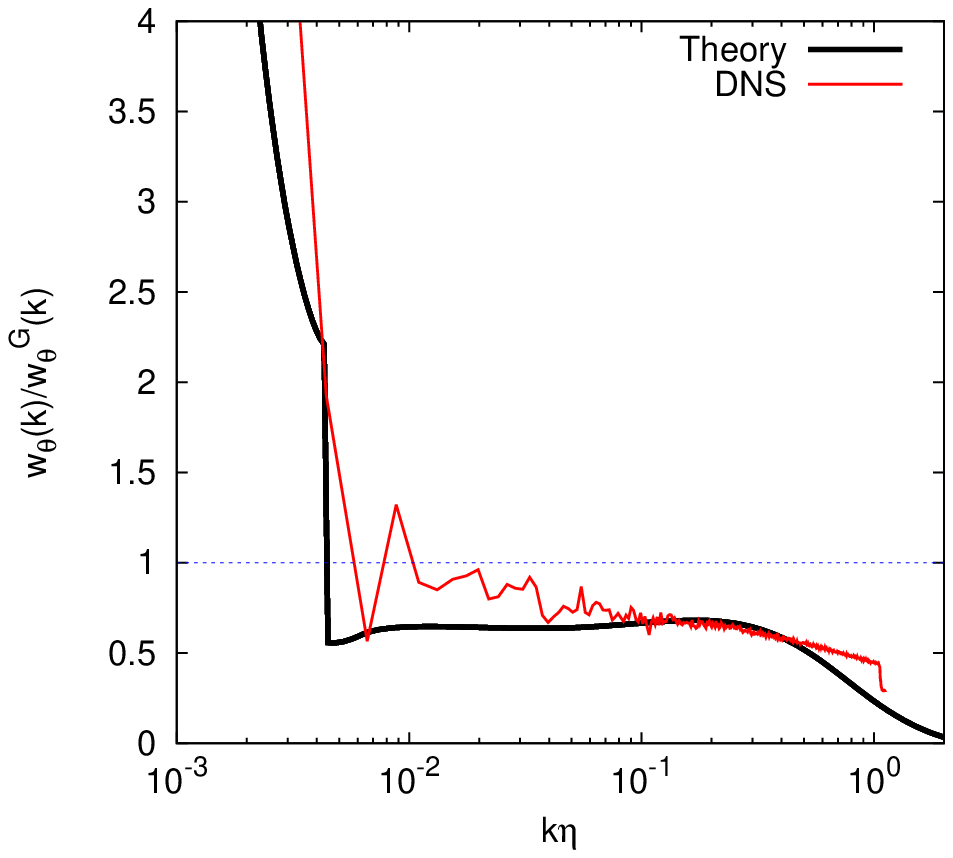}
\caption{DNS and theoretical results in isotropic turbulence at a Taylor-scale
 Reynolds number of $427$ and ${\mathcal Pr}=1$. Left: spectrum of the mean square advection term of the
 scalar equation in isotropic turbulence. Also shown are the Gaussian spectra (thin lines). Right: ratio of the spectra to the Gaussian spectra. 
\label{Fig2} }      
\end{figure}

In this section, we confirm the reduction of mean-square advection
in DNS data,\cite{Herring1992} and compare the results with closure predictions. We have computed the scalar spectrum and energy spectrum by closure theory as described in
the previous section and the parameters have been chosen to match the
DNS as closely as possible.
The DNS database used is 
from high resolution
($1024^3$ gridpoint) pseudospectral direct numerical simulations 
of a passive scalar advected by isotropic turbulence;\cite{Watanabe2004}. The force terms for the velocity and scalar fluctuations are random-Gaussian and delta-correlated in time (and solenoidal for the velocity forcing), acting in the wave-number range $1\le k \le 2$. In these simulations the Reynolds number based on the Taylor microscale is equal to $427$ and the Prandtl number
${\mathcal Pr}=\nu/\alpha=1$.
The resolution is higher than that used in the simulations
of both Herring and M\'etais\cite{Herring1992}
and Chen {\em et al.}\cite{Chen1989}. 

Using DNS data, $w_\theta(k)$ can be determined 
from Eqs. (\ref{W defn}) and (\ref{defwt}). The contribution
$w_\theta^G(k)$ is obtained by randomizing the phases of the
Fourier amplitudes of $\theta(\bm x,t)$;
this randomization will yield scalar fields with statistically 
independent Fourier 
amplitudes without changing the wavenumber spectra.
This independence, not the probability density function itself, 
is the key property for us. The fields are therefore random-phase fields and the amplitude statistics are not necessarily Gaussian.

Figure \ref{Fig1} compares the scalar variance spectra in DNS
and the closure computations at $R_\lambda=427$ and ${\mathcal Pr}=1$. The inset shows the energy spectra. The wavenumber of these results is normalized 
by the Kolmogorov scale, which is equal to the Batchelor scale 
for unit Prandtl number. Good agreement is observed between the DNS results and the EDQNM results. A Corrsin-Obukhov inertial range for the scalar spectrum and a Kolmogorov inertial range for the energy spectrum, both approximately proportional to $k^{-5/3}$ are clearly observed. 

In Figure \ref{Fig2}, left, the spectrum of the advection term $w_\theta(k)$ is shown, as well as its Gaussian estimate. These spectra, for both closure and DNS, display an increasing trend in the inertial range and peak around the Batchelor scale. The peak of $w_\theta(k)$ is smaller then the Gaussian value, which indicates a reduction of mean-square advection. This reduction is more clearly observed in Figure \ref{Fig2}, right, in which we display the measure of the departure from Gaussianity, 
the ratio $w_\theta(k)/w_\theta^G(k)$ \cite{Herring1992}; the analogous quantity for the
velocity field was introduced 
by Kraichnan and Panda.\cite{Kraichnan1988} The ratio departs noticeably from the
Gaussian values over the entire wavenumber range, and
a significant depression of the $w_{\theta}(k)$ compared to the
Gaussian value is observed at scales larger than the forcing scales. 
The region where $w_\theta(k)/w_\theta^G(k)<1$ extends over the entire inertial-convective range. 
These general trends, including the observation that 
$w_\theta(k)/w_\theta^G(k)>1$ at large scales,
are consistent with previous observations.\cite{Chen1989,Herring1992}. The results in Figures \ref{Fig1} and \ref{Fig2} show that the closure yields results in good agreement with the DNS results.

The ratio of the measured variance to the value assuming independence of
the Fourier amplitudes,
\begin{equation}
\rho_\theta=\frac{\langle ({\bm u}\cdot\nabla\theta)^2 \rangle}{\langle 
|{\bm u}|^2\rangle\langle|\nabla\theta|^2\rangle}
=
\frac{\displaystyle\int^{\infty}_0 w_\theta(k)dk}{\displaystyle\int^{\infty}_0 w^G_\theta(k)dk} 
\label{ratio}
\end{equation}
is also of interest. Figure \ref{Fig2} (left) shows that the spectrum
$w_\theta(k)$ is an increasing function of the wavenumber, consequently
its integral is dominated by the small scales, where the variance is reduced.
The DNS value for $\rho_\theta$
is 0.41 and the closure value is 0.54. 
These values are consistent
with the previous reported results: Herring and M\'etais\cite{Herring1992} quotes a
value for $\rho_\theta$ of about 0.5 in the scalar case, and Kraichnan and Panda \cite{Kraichnan1988} 
reported the value 0.57 for the comparable ratio of the mean-square nonlinearity.
We conclude that the effect we investigate is observed in DNS and closure
and is of comparable magnitude in both.

\begin{figure}
\includegraphics[width=.5\textwidth]{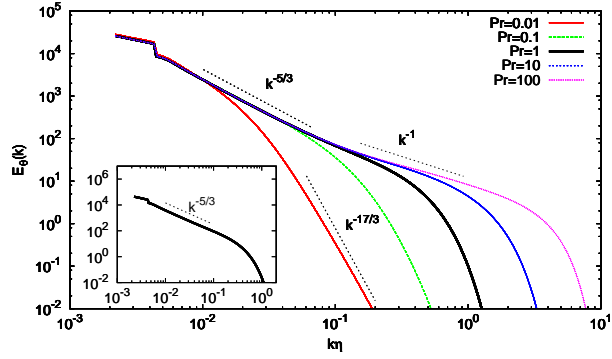}
\caption{Closure results for the spectrum of the scalar variance at a
  Taylor-scale Reynolds number of $427$ and ${\mathcal
  Pr}=0.01,..,100$. Inset: energy spectrum.}
\label{Fig3}       
\end{figure}

\begin{figure}
\includegraphics[width=.5\textwidth]{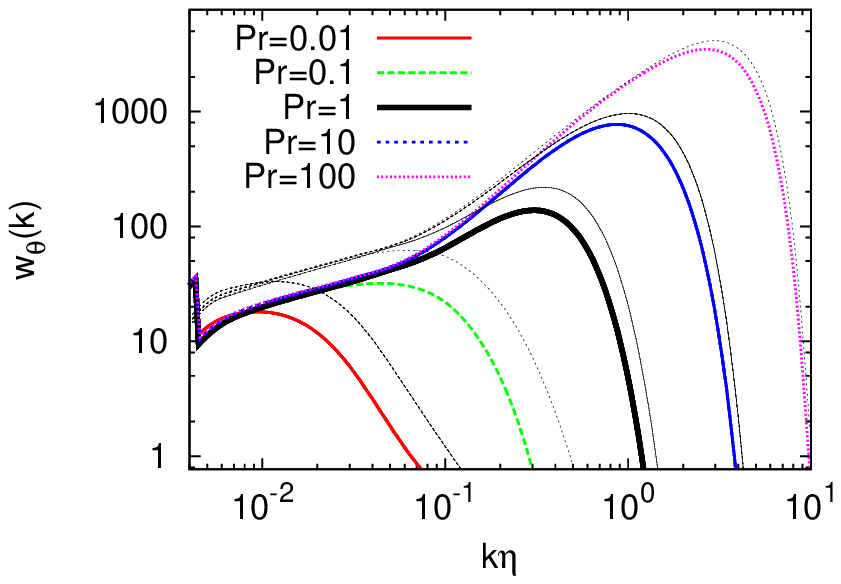}\includegraphics[width=.5\textwidth]{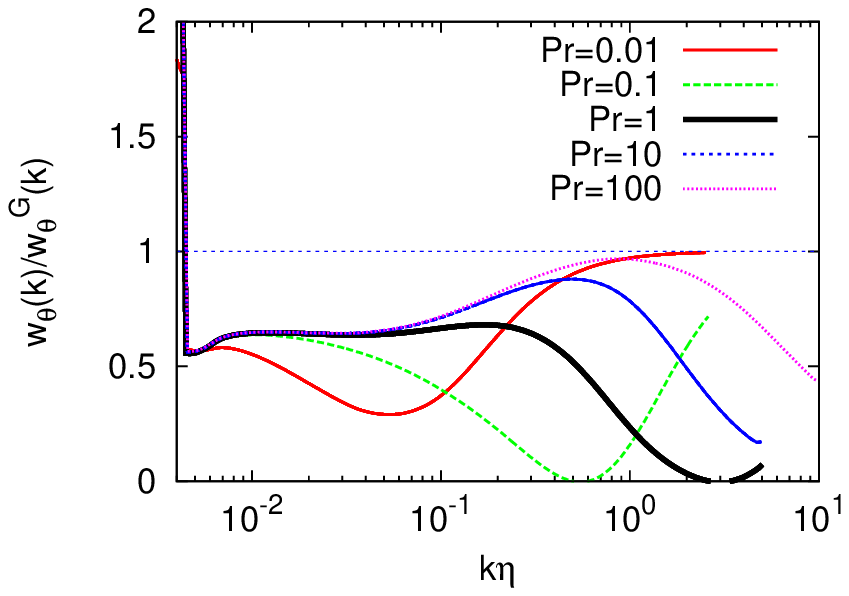}
\caption{Closure results for the spectrum of the mean square advection term at a
  Taylor-scale Reynolds number of $427$ and ${\mathcal
  Pr}=0.01,..,100$. Also shown are the Gaussian spectra (thin lines). Right: ratio of the spectra to the Gaussian spectra.}
\label{Fig4}       
\end{figure}



\begin{figure}
\includegraphics[width=.5\textwidth]{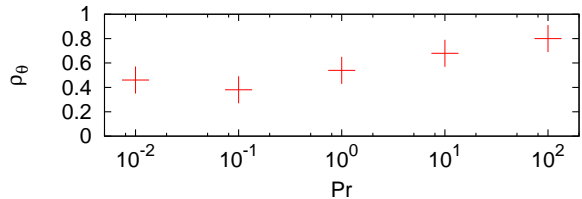}
\caption{The mean square advection term of the
  scalar equation in isotropic turbulence compared to its Gaussian value as a function of the Prandtl number at a Taylor-scale Reynolds number of  $427$.
\label{FigPr}       }
\end{figure}

In problems involving passive scalars, the dependence on Prandtl
number is always of interest.
We investigate the effect of the Prandtl number on the reduction of
mean-square advection by
varying the Prandtl number from $0.01$ to $100$ at a fixed
Reynolds number $R_\lambda=427$. There is no DNS data available for these cases, in particular for the high Prandtl number cases,
so we limit the discussion to closure predictions.
Figures \ref{Fig3} and \ref{Fig4} show the closure results. In Figure \ref{Fig3} we show the scalar spectrum for five different Prandtl numbers. At low Prandtl numbers the $k^{-17/3}$ spectrum is observed and at large ${\mathcal Pr}$ we observe a $k^{-1}$ spectrum \cite{Batchelor1959-2,Batchelor1959}.

Figure \ref{Fig4} (left) shows the spectrum of the advection term. For all ${\mathcal Pr}$, this spectrum is an increasing function of the wavenumber. At the highest value of ${\mathcal Pr}$, the spectrum seems to approach its Gaussian estimate. Figure \ref{Fig4} (right) shows $w_{\theta}(k)/w_{\theta}^G(k)$. It is clearly observed that the spectrum is under its Gaussian value for all scales, except the forced scales, but the precise behavior seems to depend strongly on the Prandtl number.  In the inertial-convective range the depletion is approximately constant with respect to wavenumber. The effect is weaker in the Batchelor range.

The numerical values of $\rho_\theta$ are displayed in Figure \ref{FigPr}. The value ranges from a minimum of $\rho_\theta=0.38$ at ${\mathcal Pr}=0.1$ to a maximum of $\rho_\theta=0.8$ at ${\mathcal Pr}=100$. This change is non-negligible, but the trend is
rather weak if we consider that ${\mathcal Pr}$ changes over four orders of magnitude in our simulations. The reduction of advection seems thus an effect which is persistent, but becomes weaker for high values of the Prandtl number. Its amount is mainly determined by the precise behavior of the cumulant-spectrum around the scale where the spectrum $w_{\theta}(k)$ peaks.


\section{Discussion: mechanisms of the suppression of advection \label{sec:Discussion}}

\begin{figure}
\includegraphics[width=.35\textwidth]{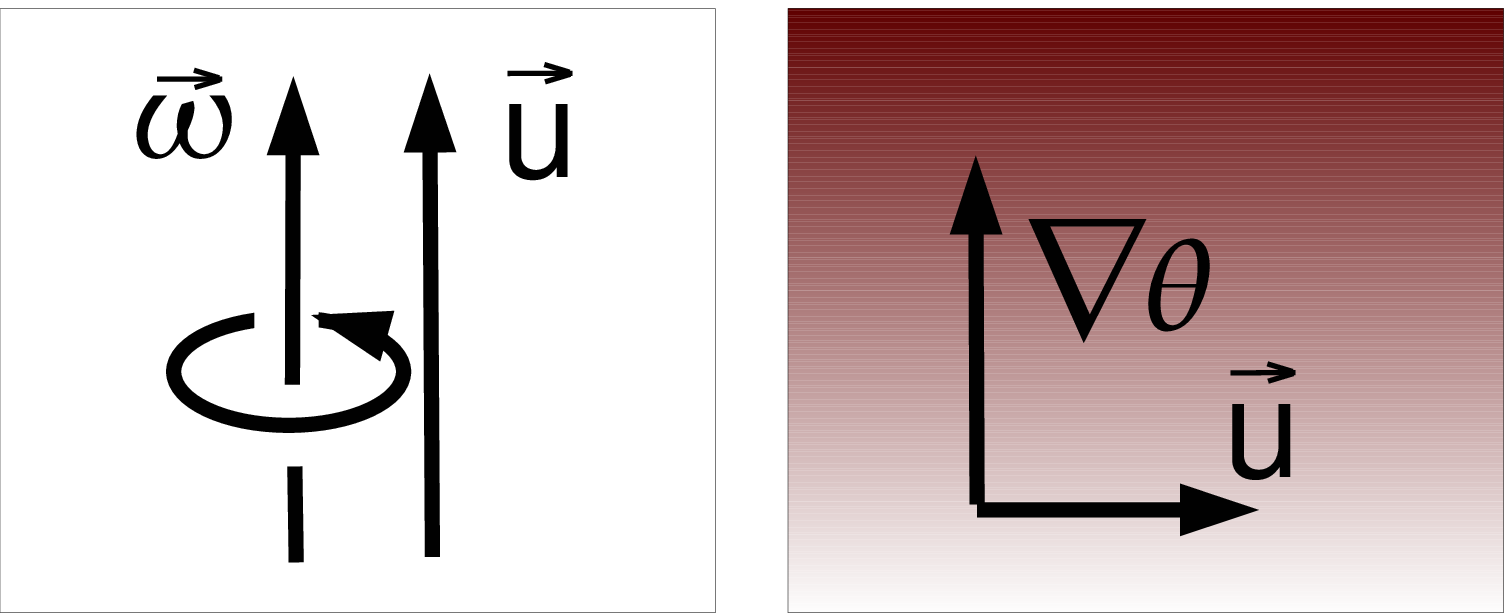}\\
\includegraphics[width=.35\textwidth]{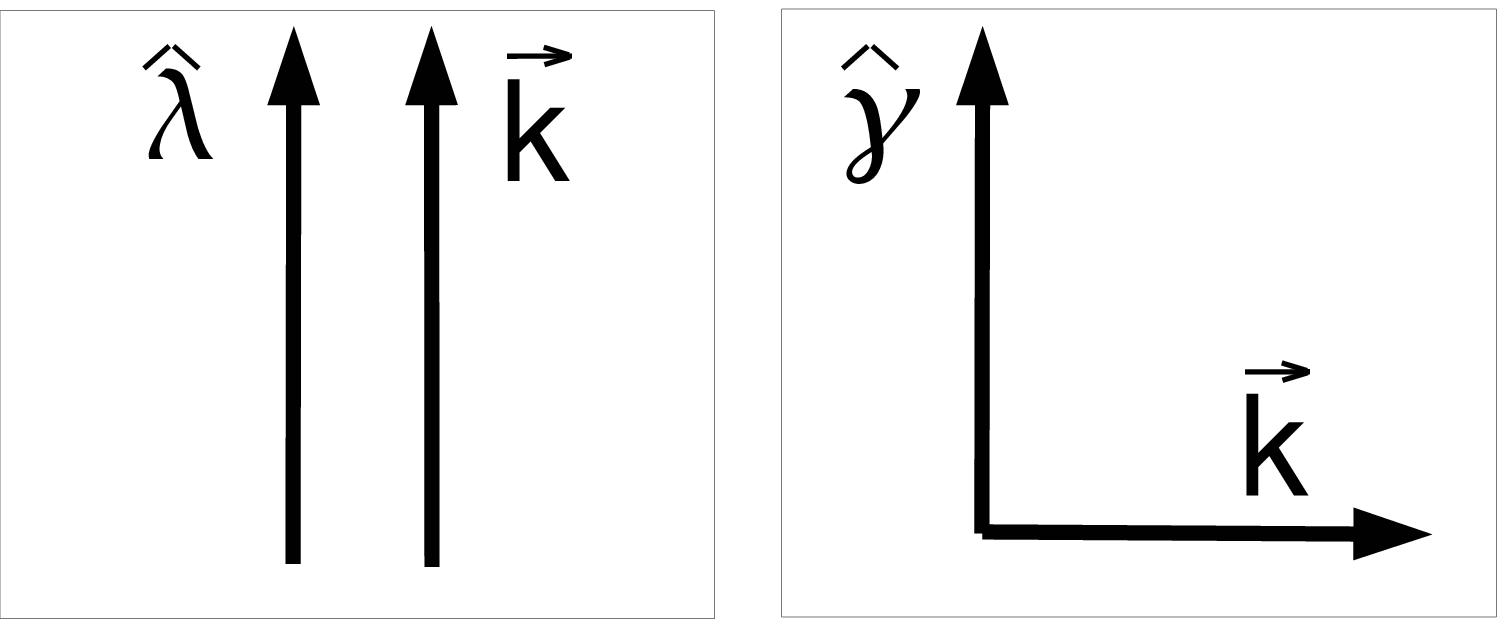}

\caption{Left, top: velocity Beltramization; bottom: depression of nonlinearity through alignment of the Lamb-vector and the wavevector. Right: depletion of advection in physical and Fourier space. The scalar flux vector is defined as $\bm \gamma=\bm u \theta$.}
\label{FigBeltrami}       
\end{figure}

The analysis of the variance of the nonlinear term in the
Navier-Stokes equations by Chen \emph{et al.}\cite{Chen1989} was motivated
in part by 
the suggestion of Levich and Tsinober \cite{Levich1983} of
the possibility of {\em Beltramization}, the preferential alignment of velocity and vorticity in
turbulence. 
Since the nonlinear term can be written as
\begin{equation}\label{NSk}
\nabla^{-2}\nabla\times\nabla\times
(\bm\omega(\bm x)\times \bm u(\bm x)),
\end{equation}
with $\bm \omega=\nabla\times\bm u$ the vorticity, 
this alignment will obviously reduce the
magnitude of the nonlinear term, and hence will also 
reduce the intensity of its fluctuations, which is consistent with the
observed depression of nonlinearity.

Another mechanism consistent with depression of nonlinearity was
identified by Kraichnan and Panda,\cite{Kraichnan1988}
who noted that the nonlinearity also vanishes if the
Lamb vector $\bm \lambda(\bm x)\equiv \bm \omega(\bm x)\times \bm u(\bm x)$
is a potential field
($\bm \lambda(\bm x)=\nabla\Phi(\bm x)$), so
that it lies 
in the null-space of the double curl operator in Eq. (\ref{NSk}). 
These two 
possibilities are illustrated in Figure \ref{FigBeltrami} (left). 
Both possibilities can contribute to the depression of nonlinearity in 
turbulent flows.

The situation is much simpler for scalar advection.
For the passive scalar, the equivalent of Beltramization
would be the identical vanishing of the scalar flux vector $\bm \gamma=\bm u \theta$;
this trivial case can be ignored.  
The only non-trivial way to reduce the 
advection term is for the scalar flux vector
to be divergence-free, so that
\begin{equation}
\nabla\cdot \bm \gamma = \bm u\cdot\nabla\theta\approx 0.
\end{equation}
This corresponds to the case in which the velocity is perpendicular to 
the scalar gradient, as illustrated in Figure \ref{FigBeltrami} (right). 
It is evident that if the variance of the advection term is smaller
in passive scalar advection than in a jointly Gaussian random field,
then $\bm u$ and $\nabla\theta$ {\em must} be more likely to be orthogonal 
in passive scalar advection than in a jointly Gaussian random field.

\section{conclusions \label{sec:conclusions}}
We have shown that the closure computation of the fourth order cumulant that
enters in the depression of nonlinearity in hydrodynamic turbulence \cite{Chen1989}
can be applied to passive scalar advection. Corresponding to depression of
nonlinearity is a reduction of the variance of the advection term, 
which is connected to a tendency of the velocity vector to
align perpendicular to the scalar gradient. Study at the level of Fourier
spectra shows that the reduction of advection is approximately constant in the inertial-convective range and becomes weaker in the viscous-convective (Batchelor) range.
Closure related to the DIA gives 
satisfactory predictions in comparison to DNS data. Closure predicts that
the phenomenon persists at both low and high Prandtl numbers although there is a weak
but noticeable
tendency for the mean-square advection to return to the Gaussian value
as the Prandtl number increases.

\emph{Acknowledgments.} The authors are indebted to Toshiyuki Gotoh for discussion and for making the DNS data available. Part of the DNS data was downloaded from the CINECA database.


\begin{thebibliography}{10}
\bibitem{Footnote2}
Proudman's very favorable assessment of this idea is noteworthy, since otherwise,
his assessment of DIA was sharply critical and even dismissive.
\end{thebibliography}
\end{document}